\documentclass[%
twocolumn,
 amsmath,amssymb,
 aps,authoryear
]{aastex63}
\usepackage{graphicx}
\usepackage{ulem}
\usepackage{epstopdf}
\usepackage{amsmath}
\usepackage{amssymb}
\usepackage{mathtools}
\usepackage{mathrsfs}
\usepackage{bm}
\setcitestyle{authoryear,round} 
\usepackage{color}
\usepackage{xcolor}
\usepackage{textcomp}
\usepackage{gensymb}
\usepackage{hyperref}
\shorttitle{Kinetic-scale intermittency in Near-Sun Solar Wind Turbulence - PSP Observations}
\shortauthors{Chhiber et al.}
\newcommand{\rs}{\text{R}_\odot}

\newcommand{\di}{d_\text{i}}
\newcommand{\pmax}{p_\text{max}}
\newcommand{\de}{d_\text{e}}

\newcommand{\kmps}{\text{km/s}}          
\newcommand{\psp}{\textit{PSP}}

\newcommand{\isois}{IS\(\odot\)IS}
\begin{document}
\title{Subproton-scale Intermittency in Near-Sun Solar Wind Turbulence Observed by the Parker Solar Probe}
\author[0000-0002-7174-6948]{Rohit Chhiber}
\correspondingauthor{Rohit Chhiber}
\email{rohitc@udel.edu}
\affiliation{Department of Physics and Astronomy \& Bartol Research Institute, University of Delaware, Newark, DE, USA}
\affiliation{Heliophysics Science Division, NASA Goddard Space Flight Center, Greenbelt, MD, USA} 
\author[0000-0001-7224-6024]{William H.~Matthaeus}
\affiliation{Department of Physics and Astronomy \& Bartol Research Institute, University of Delaware, Newark, DE, USA}
%
\author[0000-0002-4625-3332]{Trevor A.~Bowen}
\affiliation{Space Sciences Laboratory, University of California Berkeley, Berkeley, CA, USA}

\author[0000-0002-1989-3596]{Stuart D.~Bale}
\affiliation{Space Sciences Laboratory, University of California Berkeley, Berkeley, CA, USA}
\affiliation{Physics Department, University of California Berkeley, Berkeley, CA, USA}

\begin{abstract}
High time-resolution solar wind magnetic field data is employed to study statistics describing intermittency near the first perihelion \((\sim 35.6~\rs)\) of the \textit{Parker Solar Probe} mission. A merged dataset employing two instruments on the FIELDS suite enables broadband estimation of higher order moments of magnetic field increments, with five orders established with reliable accuracy. The duration, cadence, and low noise level of the data 
permit evaluation of scale dependence of the observed intermittency from the inertial range to deep subproton scales. The results support multifractal scaling in the inertial range, and monofractal but non-Gaussian  scaling in the subproton range, thus clarifying suggestions based on data near Earth that had remained ambiguous due to possible interference of the terrestrial foreshock. The physics of the transition to monofractality remains unclear but we suggest that it is due to a scale-invariant population of current sheets between ion and electron inertial scales; the previous suggestion of incoherent kinetic-scale wave activity is disfavored as it presumably leads re-Gaussianization which is not observed.
\end{abstract}
\section{Introduction}\label{sec:intro}
Intermittency
is an important feature 
in the theory of fluid and plasma turbulence \citep{sreenivasan1997AnRFM,matthaeus2015ptrs}, and 
has gained increasing attention 
in the 
study of space plasmas, including the 
corona, the magnetosphere, and the solar wind \citep{ abramenko2008,chhiber2018MMS,bruno2019ESS}. In each of these venues the emergence
of localized \replaced{strong gradients}{sporadic features of the primitive fields} is a consequence of the turbulent cascade of energy.
The resulting coherent structures, of 
electric current density,  vorticity, or density, 
are likely sites of 
enhanced
kinetic dissipation,
and heating \citep[e.g.,][]{osman2012PRL}.
Therefore intermittency 
is crucial 
in terminating the 
cascade and 
heating the plasma. 
These 
coherent structures 
also compartmentalize the plasma, \added{forming boundaries associated with} distinctive flux tube ``texture'' that organizes quantities such as temperature, density, magnetic intensity,
and energetic particles \citep{borovsky2008JGR,tessein2013ApJ}.
Coherent structure forms 
in similar ways in hydrodynamics \citep{sreenivasan1997AnRFM}, magnetohydrodynamics \citep{wan2009PoP}, and plasmas \citep{burlaga1991JGR,sorriso-valvo1999GRL,matthaeus2015ptrs}, with 
important differences, particularly approaching kinetic scales. Statistics in the
plasma kinetic range  provide 
insight regarding 
physical mechanisms responsible for 
dissipation \citep[e.g.,][]{goldstein2015RSPTA,chen2016jpp,matthaeus2020},
thus addressing fundamental questions related to coronal heating and acceleration of the solar wind \citep{fox2016SSR}.

\added{Qualitatively speaking, monofractality is associated with structure that is non space-filling but lacking a preferred scale over some range (i.e., scale-invariance). In contrast, multifractality also implies non space-filling structure but with at least one preferred scale within the relevant range \citep{frisch1995book}. Such a distinction has likely implications for the preference of a system for specific classes of dissipative mechanisms. In the solar wind, and in collisionless plasmas in general,} it remains unclear whether statistics at subproton scales remain strongly intermittent and multifractal, or become monofractal \citep{kiyani2009PRL,leonardis2013prl,leonardis2016pop,wan2016PoP}, or even return to 
Gaussianity \citep{koga2007pre,wan2012ApJ,wu2013ApJ,chhiber2018MMS,roberts2020PRR}. 
These questions persist in part because of the scarcity of high time-resolution data at locations well separated from the terrestrial bow shock. Here we address these issues by employing high-resolution measurements of the magnetic field made by the \textit{Parker Solar Probe} (\psp ) in near-Sun solar wind \citep{fox2016SSR}.

\section{Theoretical Background}\label{sec:theory}
In turbulence, considerable information 
is contained in the 
statistics of fluctuations and increments of the primitive variables. These
are velocity in hydrodynamics, velocity and magnetic fluctuations in magnetohydrodynamics (MHD), density for compressible flows, and additional variables for complex fluids and plasmas. The basic second-order statistics 
include 
two-point correlations, 
their Fourier transforms, i.e., wavenumber ($k$) spectra \citep{matthaeus1982JGR},  and the second-order structure functions
\citep{burlaga1991JGR}. 
These and other relevant statistics are moments of the underlying joint probability distributions functions \citep[PDFs; e.g.,][]{Monin1971book}.
Second-order moments describe the distribution of energy over spatial scales $\ell\sim 1/k$. To describe the spatial concentration of energy in {\it intermittent} structures, we go beyond second-order statistics and consider {\it higher-order} moments of PDFs \citep[e.g.,][]{frisch1995book}. 


The original K41 similarity hypothesis \citep{kolmogorov1941DoSSR} postulates the statistical behavior of longitudinal velocity increments at spatial lag \(\bm{\ell}\), namely 
 $   \delta u_\ell 
    = \bm{\hat{\ell}} \cdot
     [ \bm{u} (\bm{x} + \bm{\ell}) - \bm{u}(\bm{x}) ]$.
K41 asserts that 
$ \delta u_\ell \sim
\epsilon^{1/3} \ell^{1/3}$ where $\epsilon$ is the total 
dissipation rate and isotropy is assumed. Thus, for an appropriate averaging operator \(\langle \dots \rangle\), all 
increment moments are determined as the structure functions
$S^{(p)} = \langle \delta u_\ell^p \rangle = C_p \epsilon^{p/3} \ell^{p/3}$, 
a form that includes the second-order law $S^{(2)} = C_2 \epsilon^{2/3} \ell^{2/3}$ and (formally) the exact third-order law $S^{(3)}= -(4/3) \epsilon \ell$ as special cases. 
The refined similarity hypothesis \citep[][K62]{kolmogorov1962JFM}
takes into account intermittency, averaging the 
local dissipation rate $\epsilon_\ell$
over a volume of linear dimension $\ell$ and introducing this as an additional
random variable. 
Incorporating the suggestion \citep{oboukhov1962JFM}
that such irregularity of dissipation changes scalings of increments with $\ell$, 
the refined 
K62 hypothesis becomes 
$\delta u_\ell = A(*) 
\epsilon_\ell ^{1/3} \ell^{1/3}$. 
Here 
$A(*)$ is a random function that depends on 
local Reynolds number, but not on $\epsilon_\ell$ or $\ell$ separately, 
and takes on a unique form at infinite Reynolds number. 
For 
moments $S^{(p)} = 
\langle \delta u_\ell^p \rangle$, the 
hypothesis implies  \begin{equation}
S^{(p)} = C_p \langle \epsilon_\ell^{p/3}\rangle \ell^{p/3} = C_p \epsilon^{p/3} \ell^{p/3 + \mu(p)}
    \label{eq:K62}
\end{equation} 
where $\mu(p)$ is a measure of the 
intermittency. We define the scaling exponent \(\zeta (p) = p/3 + \mu(p) \). 

\section{Outstanding Observational Questions}\label{sec:back_observe}
The solar wind magnetic field spectrum in the inertial range admits a power law over several decades
\citep{Coleman1966prl,matthaeus1982JGR}, although
discussion persists concerning exact spectral indices and 
anisotropy \citep[e.g.,][]{tessein2009ApJ}. PDFs of magnetic increments exhibit non-Gaussian features that are increasingly prominent at smaller lags within the inertial range \citep{sorriso-valvo1999GRL}. 
As such, it is understood that the scale-dependent kurtosis $\kappa(\ell)$ (SDK; see Section \ref{sec:intermit})
increases with decreasing $\ell$
in the inertial range, while 
higher-order exponents exhibit 
multifractal scaling \citep{frisch1995book}.
Overall, this picture is 
consistent with expectations from 
MHD \citep{carbone1995prl,politano1998EuroPhysLet}
which in turn are consistent with 
hydrodynamic scaling \citep{sreenivasan1997AnRFM}. 

The situation is less clear 
when comparing solar wind statistics in the kinetic range 
with either MHD or plasma 
studies. A major issue is the evidence that solar wind subproton-scale kurtosis {\it decreases} in the kinetic range \citep{koga2007pre,wan2012ApJ,chhiber2018MMS}. This is partially at odds with kinetic \citep{leonardis2013prl} and MHD simulation \citep{wan2012ApJ} as well as 
observations in the terrestrial magnetosheath \citep{chhiber2018MMS}. A putative decrease may be due to interference by incoherent waves from foreshock activity, or noise of instrumental or numerical origin
\citep{chian2009AnGeo,wu2013ApJ}, while a constant SDK may signify a physically relevant transition to monofractal scaling 
\citep{kiyani2009PRL,leonardis2016pop}. If incoherent plasma waves are the culprit then proximity to the terrestrial bowshock may play a role, and there are some suggestions to this effect in contrasting \textit{ACE} and \textit{Cluster} observations \citep{wan2012ApJ}. These issues are resolved below in the \psp\ observations that we present. 

\section{\psp\ Observations in near-Sun solar wind}\label{sec:data}
\begin{table*}[ht]
\centering
  \begin{tabular}{| c | c | c | c | c | c | c | c | c | c |}
    \hline 
  Time on 2018-11-06 & \(\langle V\rangle\) & \(\langle v\rangle\) & \(\langle T_\text{i}\rangle\) & \(\langle n_\text{i}\rangle\) & \(d_\text{i}\) & \(\langle B\rangle\) & \(\langle b\rangle\) & \(\langle V_\text{A}\rangle\) & \(\beta_\text{i}\)   \\ \hline    
       UTC 02:00:00 - 03:00:00 & 343 km/s & 52 km/s & \(3.8\degree\times 10^5\) K & 304 \(\text{cm}^{-3}\) & 13 km & 99 nT & 63 nT & 124 km/s &  0.4 \\ \hline
  \end{tabular}
\caption{Bulk plasma parameters. Shown are the average values of proton speed \(\langle V\rangle \equiv \langle \sqrt{V_R^2 + V_T^2 +V_N^2}\rangle\), rms velocity fluctuation \(\langle v\rangle \equiv \sqrt{\langle |\bm{V} - \langle \bm{V}\rangle|^2 \rangle}\), ion temperature \(\langle T_\text{i}\rangle\), ion density \(\langle n_\text{i} \rangle\), ion inertial scale \(d_\text{i}\), magnetic field magnitude \(\langle B\rangle \equiv \langle \sqrt{B_R^2 + B_T^2 + B_N^2}\rangle\), rms magnetic fluctuation \(\langle b\rangle\equiv \sqrt{\langle |\bm{B} - \langle \bm{B}\rangle|^2 \rangle}\), Alfv\'en speed \(\langle V_\text{A}\rangle \equiv \langle B\rangle/\sqrt{4\pi m_\text{i} \langle n_\text{i}\rangle } \), and ion beta. 
Averaging is performed over the entire interval.}\label{tab:bulk}
\end{table*}
We examine higher-order inertial and kinetic scale statistics in a region of young solar wind explored for the first time recently by \psp\ \citep{fox2016SSR}, using measurements of the magnetic field from the FIELDS instrument \citep{bale2016SSR}. We focus on a 1-hour interval near first perihelion from UTC 2018-11-06T02:00:00 to 2018-11-06T03:00:00, when \psp\ was at \(\sim 35.6~\rs\). We use the SCaM data product, which merges fluxgate and search-coil magnetometer (SCM) measurements by making use of frequency-dependent merging coefficients, thus enabling magnetic field observations from DC to 1 MHz with an optimal signal-to-noise ratio \citep{bowen2020JGR}. Solar Probe Cup (SPC) data from the SWEAP instrument \citep{kasper2016SSR,case2020ApJS} provide estimates of bulk plasma properties.

\begin{figure}
    \centering
    \includegraphics[width=.47\textwidth]{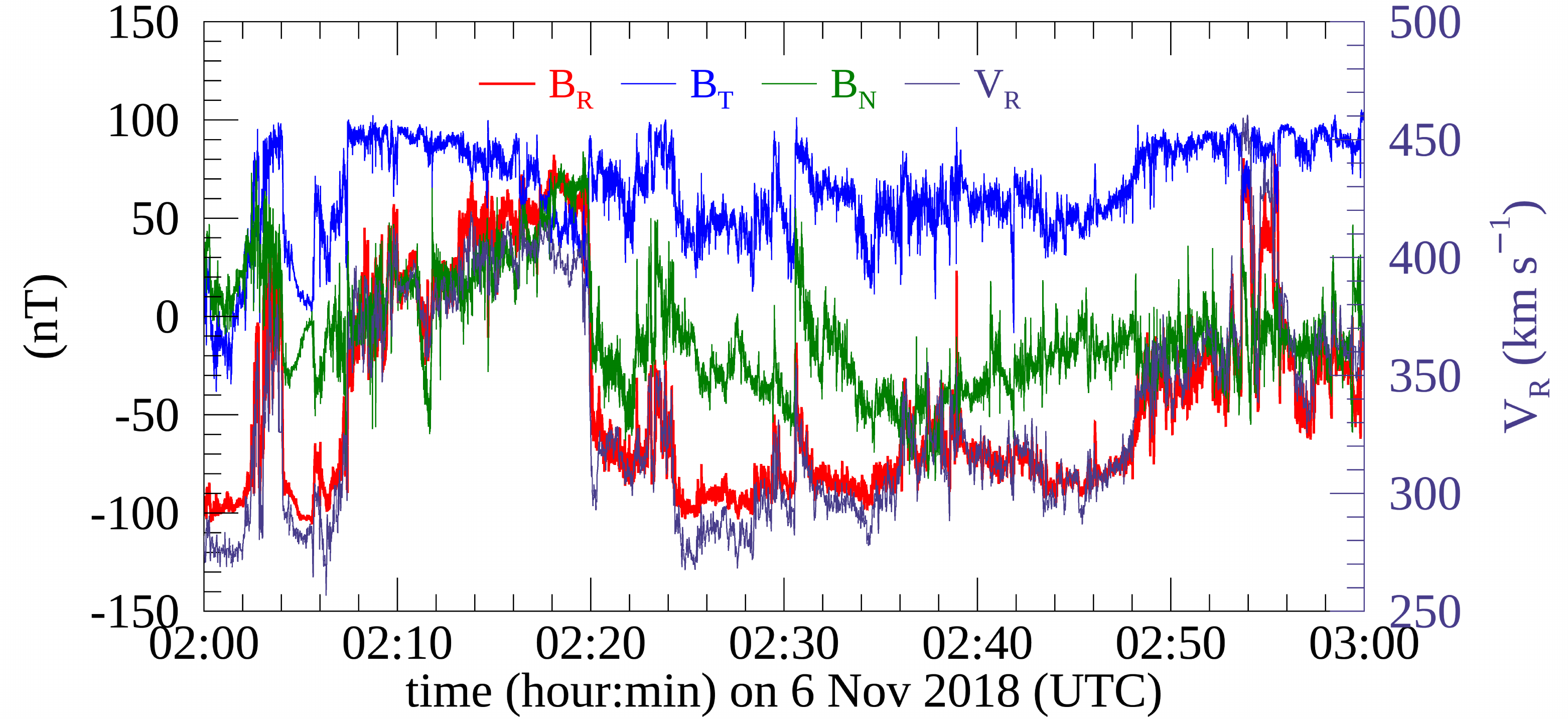}
    \includegraphics[width=.47\textwidth]{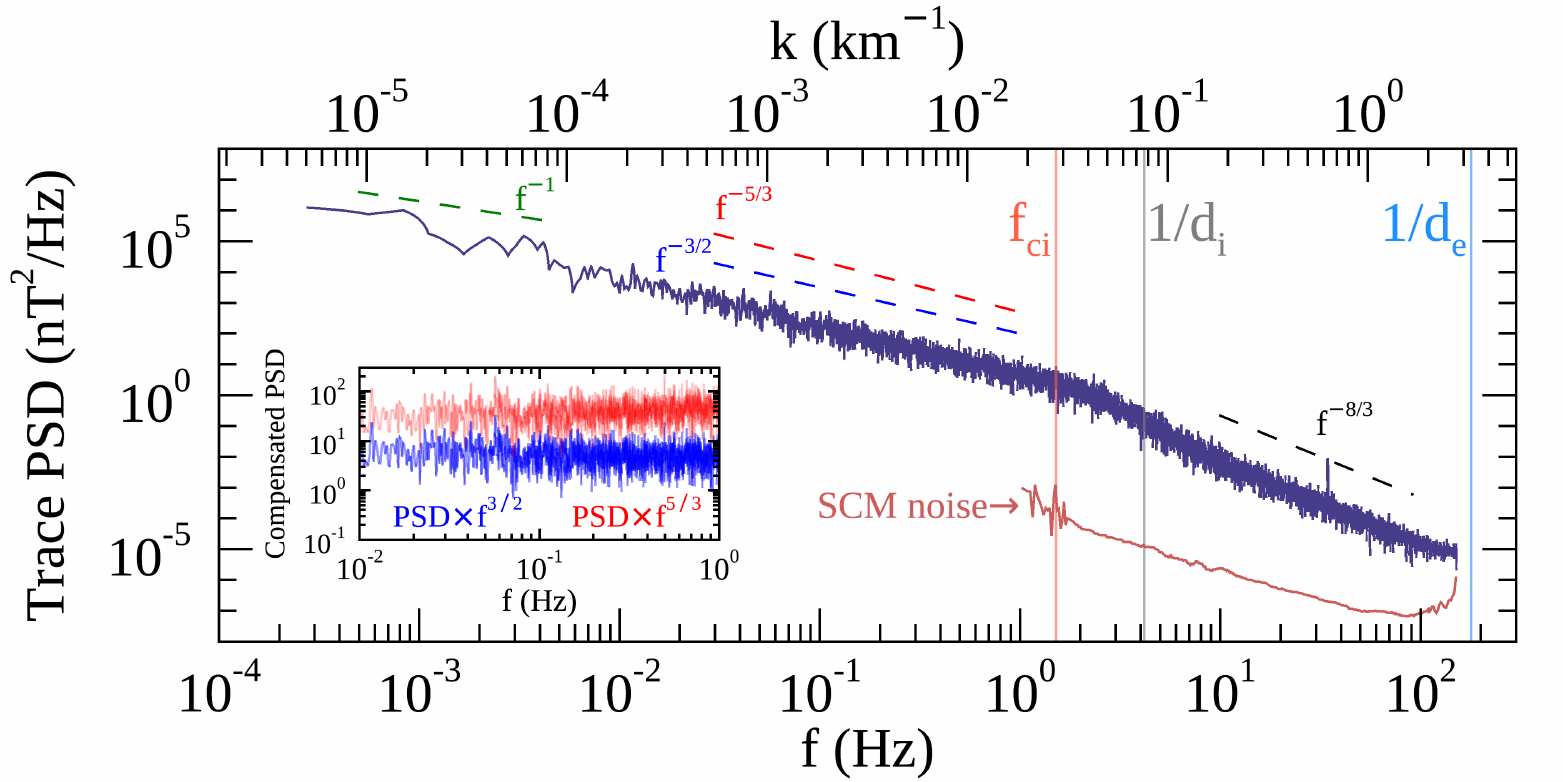}
    \caption{\textit{Top}: Time series of heliocentric RTN components of magnetic field, and radial ion velocity. \textit{Bottom}: Trace magnetic field power spectral density (PSD) \(\times\ 1/3\) (dark blue). Instrumental noise floor of SCM (brown). Inset shows compensated PSD.
    Vertical lines mark ion gyrofrequency \(f_{ci}\), and inverse of ion and electron inertial lengths (\(1/\di\) and \(1/\de\)) on wavenumber axis. Equal ion and electron densities are assumed to compute \(\de\).}
    \label{fig:tser}
\end{figure}

For the interval used here, the SCaM data-set is resampled to 0.0034 s time cadence. Time series of heliocentric \(RTN\) components \citep{franz2002pss} of the magnetic field are shown in Figure \ref{fig:tser}. SPC measurements of ion density, velocity, and thermal speed are resampled to 1 s resolution and cleaned using a time-domain Hampel filter \citep[e.g.,][]{pearson2002hampel}. The general properties of the plasma during the interval are listed in Table \ref{tab:bulk}. The radial velocity \(V_R\) during this interval indicates a slow wind, with \(V_R \lesssim 450~\kmps\). Three prominent reversals, or switchbacks \citep{DudokDeWit2020ApJS}, of the radial magnetic field are present (see Footnote \ref{ftnt:stationarity}). A high degree of correlation, or Alfv\'enicity, of velocity and magnetic field is observed \citep[][]{kasper2019Nat,chen2020ApJS}.

The correlation time \citep{matthaeus1982JGR} is \(\sim 450\) s, corresponding
to a correlation length of \(\sim 1.5\times 10^5\) km, using Taylor's frozen-in approximation \citep{taylor1938ProcRSL} with a mean speed of \(340\) km/s. Taylor's hypothesis has reasonable validity during the first \psp\ orbit \citep[][]{chhiber2019psp2,chen2020ApJS}; here it is reaffirmed by noting from Table \ref{tab:bulk} that \(\langle v\rangle /\langle V\rangle\sim 0.15\) and \(\langle V_\text{A}\rangle /\langle V\rangle\sim 0.36\).

Figure \ref{fig:tser} also shows the average power spectral density of the \(RTN\) magnetic field components. Similar spectra from \psp\ have been reported previously \citep[e.g.,][]{chen2020ApJS}. We find an inertial range that extends more than two decades in wavenumber; above the ion gyrofrequency the spectrum steepens to a \(\sim -8/3\) power law \citep[e.g.,][]{goldstein2015RSPTA}. Crucial for this work is the signal-to-noise ratio \((S/N)\) at high (kinetic range) frequencies, where the relevant instrumental noise floor is that of the SCM \citep{bowen2020JGR}, shown in Figure \ref{fig:tser}. Clearly \(S/N \ge 100\) up to 100 Hz, and remains \(\ge 5\) 
up to the highest available frequency, providing 
a measure of confidence that these 
measurements are unaffected by instrumental noise.

\section{Intermittency Observed by \psp}\label{sec:intermit}

\begin{figure}
    \centering
    \includegraphics[width=.44\textwidth]{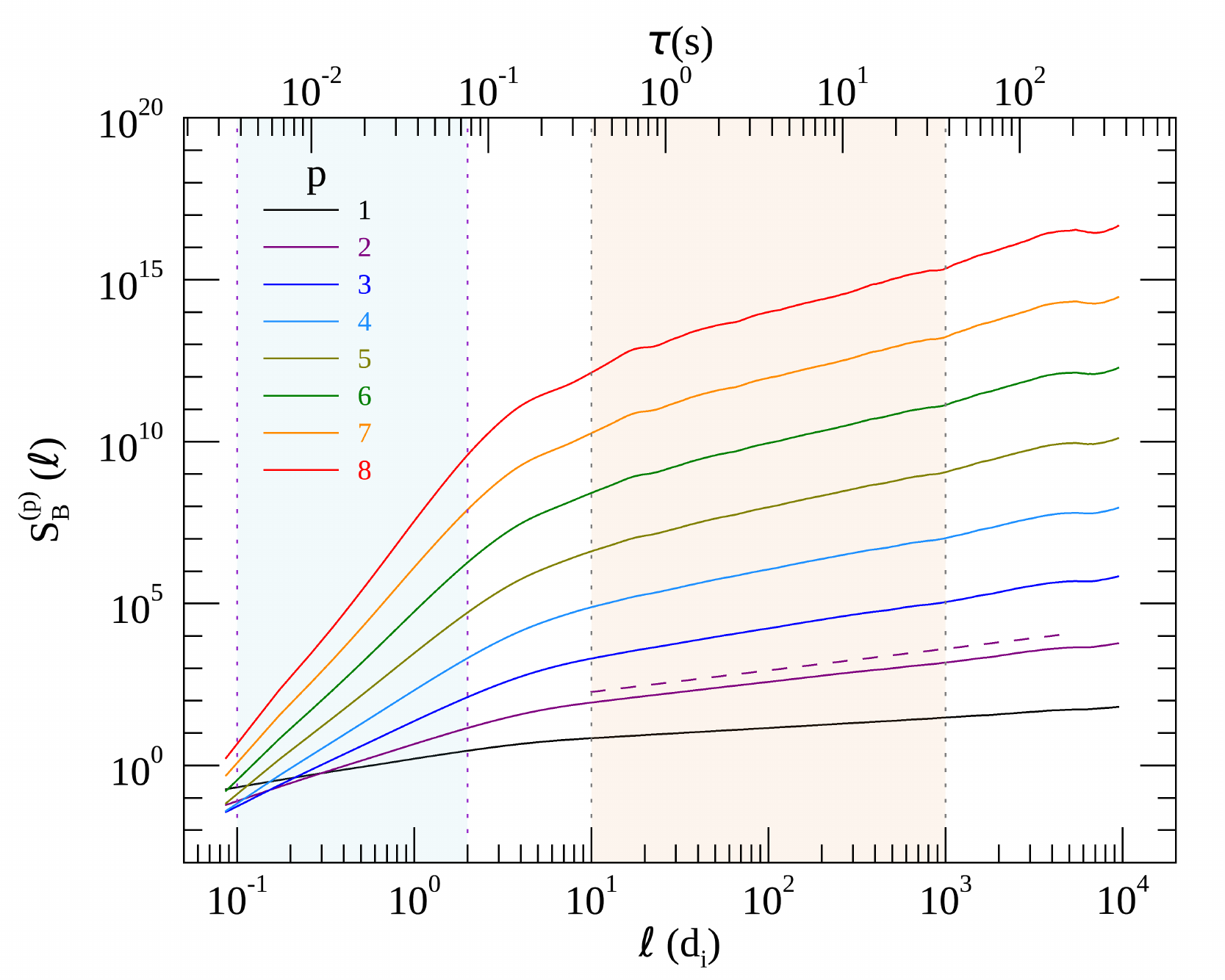}
      \includegraphics[width=.44\textwidth]{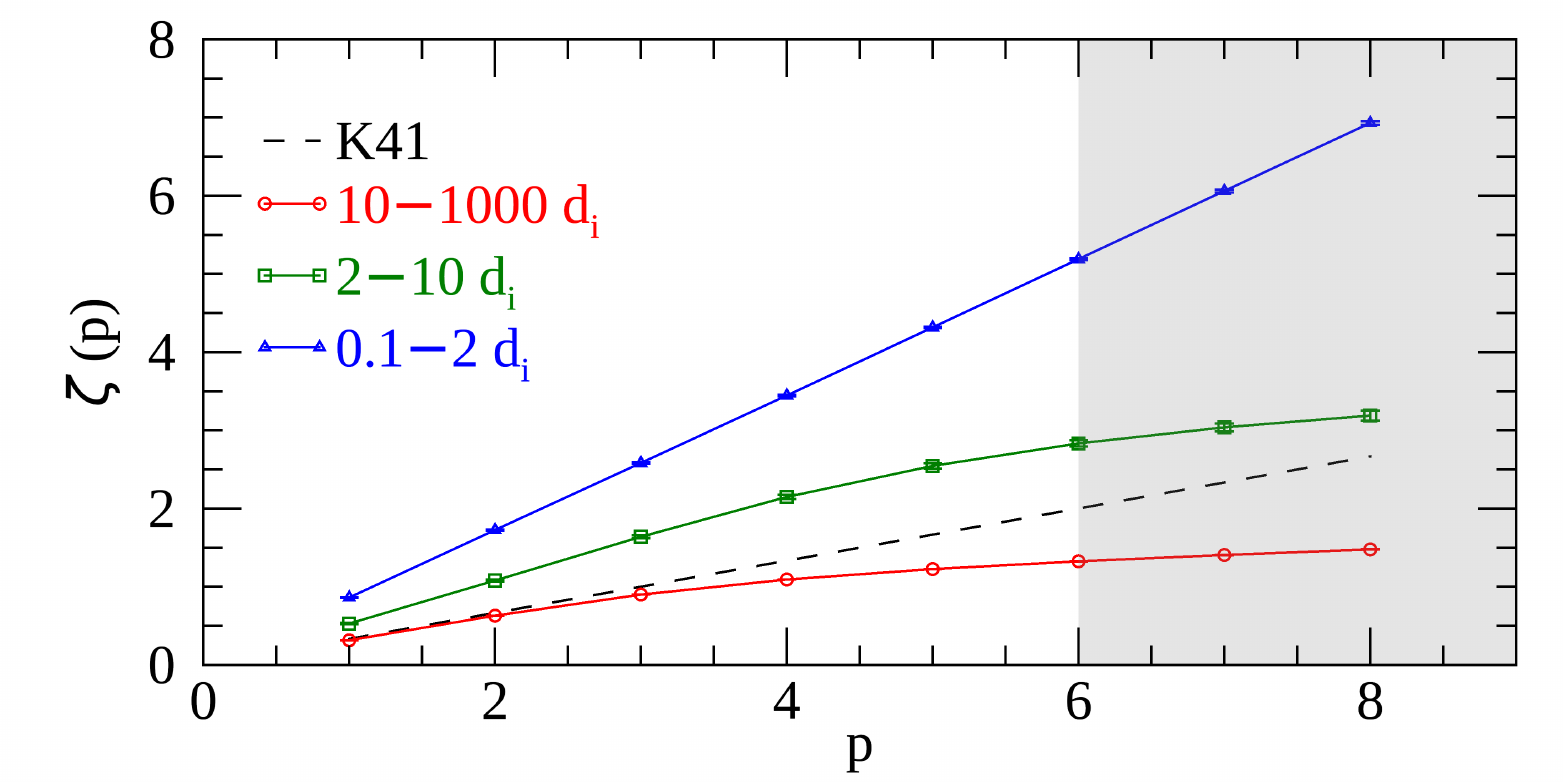}
    \includegraphics[width=.44\textwidth]{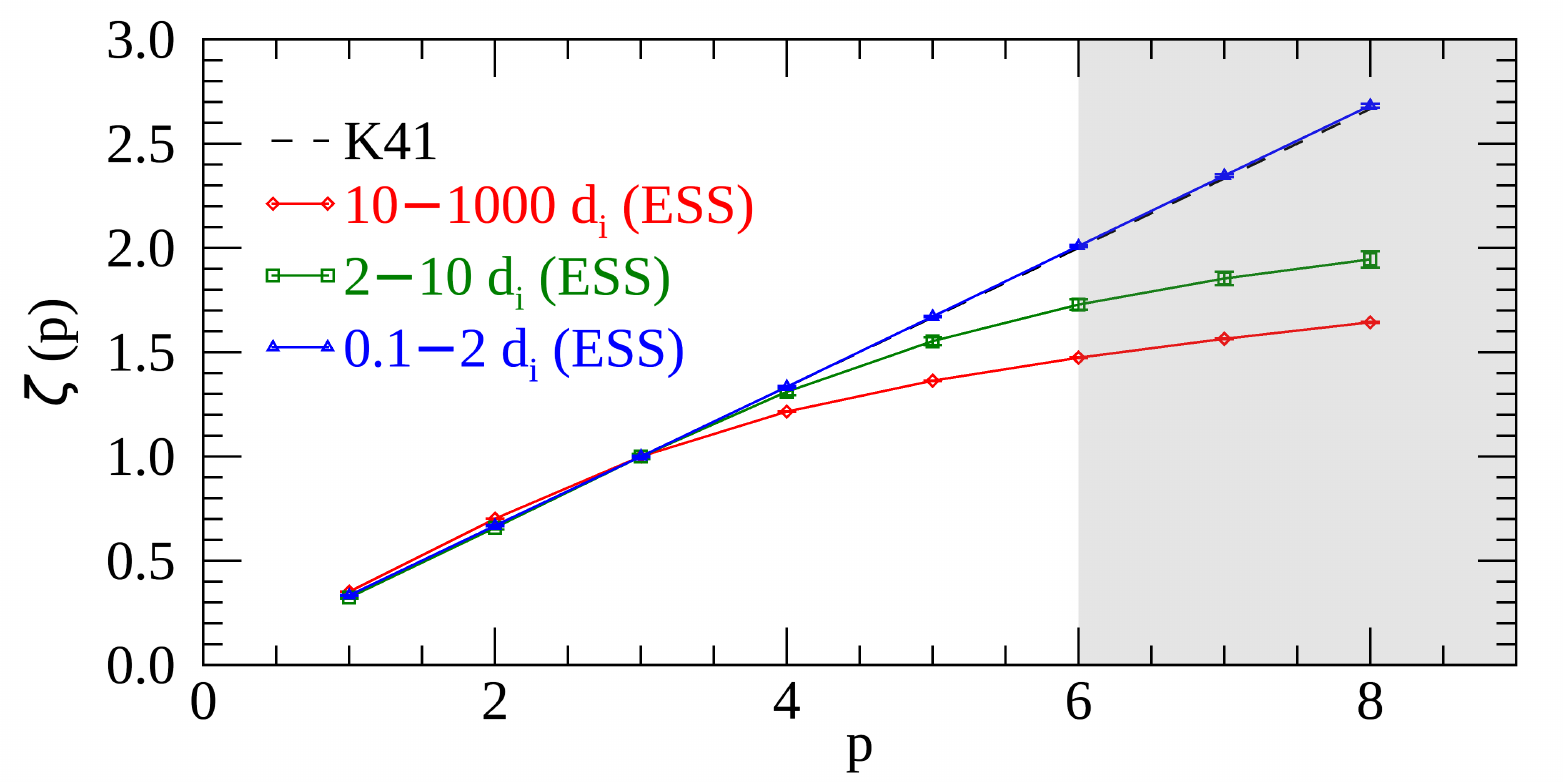}
    \caption{\textit{Top}: Structure functions for \(\delta B\)
    (Equation \eqref{eq:struc})
    {\it vs.} 
    temporal (\(\tau\)) and spatial (\(\ell\)) lags. A reference \(\ell^{2/3}\) curve (dashed, purple) is shown.
    Shaded region (cream) \(\ell = 10 - 10^3~\di\) demarcates inertial-range. Shaded region (blue) \(\ell = 0.1-2~\di\) demarcates kinetic-range. \textit{Middle}: 
    Scaling exponents \(\zeta (p)\)
    {\it vs.} $p$ for inertial, 
    kinetic, and intermediate ranges. Dashed line: K41 prediction \(\zeta (p) = p/3\). Moments not determined with reliable accuracy: grey-shaded region. \(1 \sigma\) uncertainty estimates for straight line fits to determine \(\zeta (p)\) are shown, but are generally smaller than the symbols. 
    \textit{Bottom}: Same as middle, but using ESS; scaling exponents for each range of lags are divided by \(\zeta (3)\) for the respective range.
    Kinetic-range curve (blue triangles) overlaps the K41 curve.}
    \label{fig:struct}
\end{figure}

We define increments of magnetic-field components at time \(t\) as 
\begin{equation}
    \delta B_i(t,\tau) = B_i(t+\tau) - B_i(t),\label{eq:inc2}
\end{equation}
where \(i \in \{R,T,N\}\) and \(\tau\) is a temporal lag. To convert temporal lags to spatial lags we use the Taylor approximation, wherein the spatial lag corresponding to \(\tau\) is \(\ell = \langle V_R\rangle \tau\) \citep[see][]{chhiber2020ApJS}, with mean radial solar-wind speed \(\langle V_R\rangle \sim 335~\kmps\) here. In this way we obtain spatial increments \(\delta B_i (t,\ell)\) using Equation \eqref{eq:inc2}. The magnitude of the vector magnetic increment is then \(\delta B(t,\ell) \equiv (\delta B_R^2 + \delta B_T^2 + \delta B_N^2)^{1/2}\).

The $p$-th order structure functions of \(\delta B\) are
\begin{equation}
    S^{(p)}_B (\ell) = \langle [\delta B(t,\ell)]^p \rangle_T,
    \label{eq:struc}
\end{equation}
where the \(\langle \dots \rangle\) refers to averaging over the time interval \(T \gg \tau \). Similarly, for 
for each component \({B}_i\),
\begin{equation}
    S^{(p)}_{B_i} (\ell) = \langle [\delta B_i(t,\ell)]^p \rangle_T.
    \label{eq:struc_comp}
\end{equation}
The accuracy of computed
higher-order moments is affected by 
sample size; a rule of thumb 
is that the highest order that can be computed reliably is 
\(\pmax=\log N-1\), where \(N\) is the number of samples \citep{dudokdewit2013SSR}. With \(N\sim 1.1\times 10^6\) for the present interval we get \(\pmax = 5\). Statistics of higher order than this are interpreted with some reservation. 

The top panel of Figure \ref{fig:struct} shows \(S^{(p)}_B (\ell)\) for \(p\) ranging from 1 to 8, and spatial lags \(\ell\) ranging from \(\sim 0.1~\di\), deep within the kinetic range, to  \(10^4~\di\), 
close to the energy-containing scales (the correlation length is \(\sim 1.1\times 10^4~\di\)). 
The slopes of \(S^{(p)}\) {\it vs.} \( \ell \) are larger at kinetic scales, indicating the presence of relatively stronger gradients. 
Structure functions for individual components (Equation \eqref{eq:struc_comp}) are very similar (not shown).

Next we investigate the slopes of the structure functions in greater detail. For Gaussian and non-intermittent statistics
consistent with K41 (see \S \ref{sec:theory})
one expects 
\(S^{(p)}(\ell) \propto \ell^{\zeta (p)}\) with \(\zeta (p) = p/3\). 
Figure \ref{fig:struct} (middle panel) shows the scaling exponents \(\zeta (p)\) vs \(p\), computed separately for the inertial \((10 - 10^3~\di )\) and kinetic \((0.1 - 2~\di )\) ranges, as well as an intermediate range \((2 - 10~\di)\). The exponents are computed by using chi-squared error minimization to fit straight lines to \(\ln{S^{(p)}}\) vs \(\ln{\ell}\). Inertial range exponents (red circles) begin to diverge from the K41 curve beyond \(p=3\), with higher orders showing larger departures, indicating strong intermittency with multifractal statistics (see Equation \eqref{eq:K62}). The kinetic-range curve (blue diamonds) also lies far from the K41 prediction, but is rather close to a straight line, suggesting monofractal and scale-similar but non-Gaussian statistics. These results are consistent with analyses of near-Earth solar wind turbulence based on \textit{Cluster} measurements \citep{kiyani2009PRL,alberti2019Entropy}. Exponents for the intermediate range show a transition from inertial to kinetic range behavior.

The bottom panel of Figure \ref{fig:struct} employs the Extended Self Similarity (ESS) hypothesis \citep{benzi1993PRE}, which posits that scalings of structure functions at each order are related to that of other orders. In particular the scaling of \(S^{(p)}(\ell)\) with order $p>3$ may relate better to the behavior of \(S^{(3)}(\ell )\) than to the lag $\ell$ itself. Accordingly we proceed by dividing \(\zeta (p)\) for the different lag ranges by \(\zeta (3)\) for the respective range. This rescaling does not affect the inertial range result significantly. Remarkably, the kinetic-scale exponents collapse almost perfectly to the K41 line. As far as we are aware, this has not been previously demonstrated for magnetic fluctuations in the solar wind. Similar use of ESS has been applied in kinetic simulations \citep{wan2016PoP,leonardis2016pop} and to solar wind density fluctuations at subproton scales \citep{chen2014ApJ}. The intermediate range once again exhibits transitional behavior.\footnote{The present application of ESS is somewhat at variance with the original usage \citep{benzi1993PRE}, where the significance of \(\zeta(3)\) derives from its correspondence to the third-order energy transfer law. This connection is lost in a plasma because the energy flux involves \textit{mixed} third-order structure functions of both magnetic and velocity fields \citep{politano1998PRE}; in this sense ESS should properly be based on mixed correlators \citep{politano1998EuroPhysLet}.  However, including velocity statistics here is not an option since \psp\ plasma data are not available at sufficiently high cadence to probe the kinetic range \citep{case2020ApJS}. Nevertheless, the use of ESS as we have implemented it clearly organizes the data in a revealing way.}

\begin{figure}
    \centering
    \includegraphics[width=.47\textwidth]{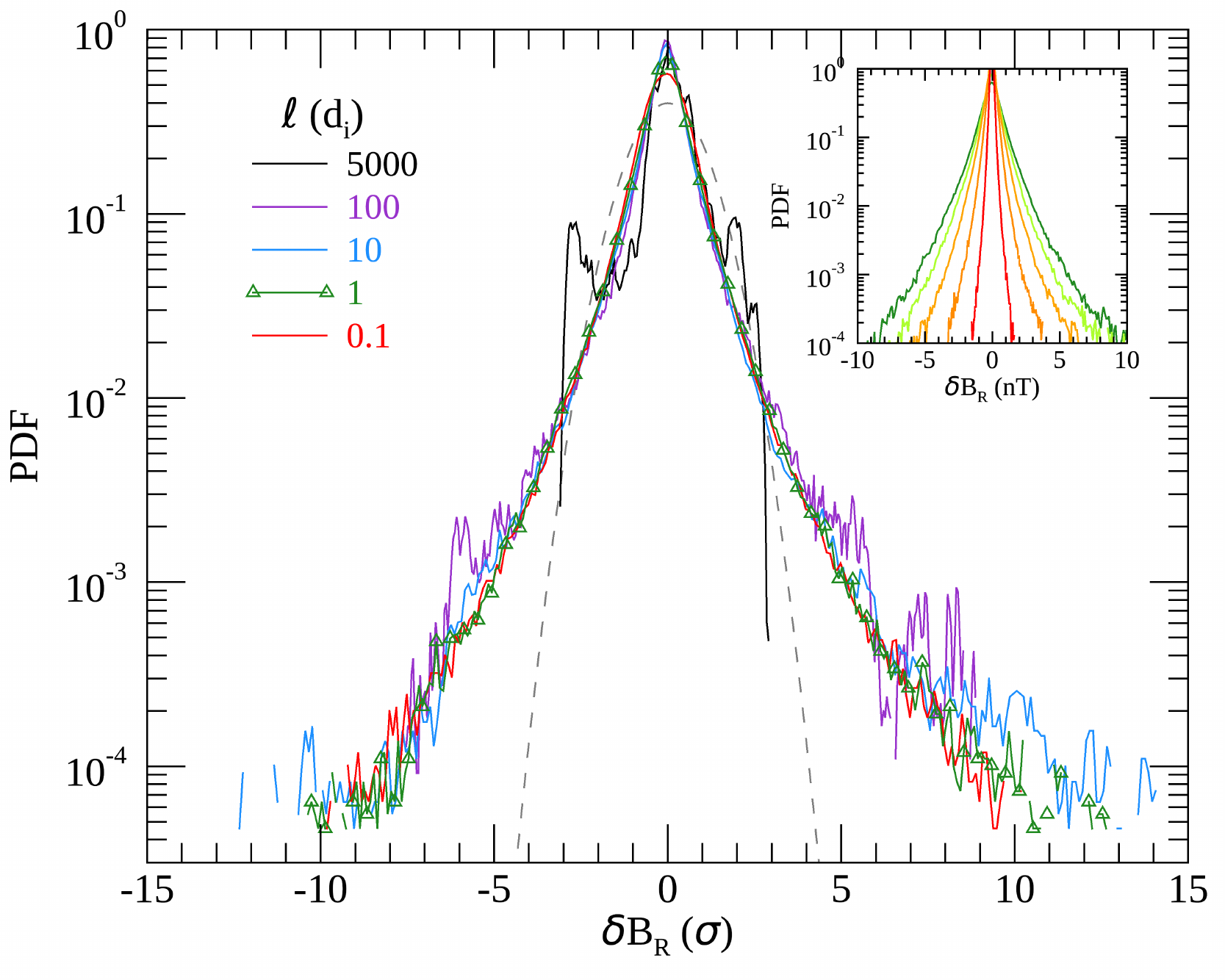}
    \caption{PDFs of \(\delta B_R\) normalized by their standard deviation 
    $\sigma$. Gaussian PDF shown for reference (dashed line). 
    \textit{Inset}: PDFs of unnormalized \(\delta B_R\); lags from \(1~\di\) (outermost curve, green) to \(0.1~\di\) (innermost, red). Collapse of green and red curves after rescaling is evident in the main graphic. All PDFs include bins with population $\ge 5$.}
    
    \label{fig:pdf}
\end{figure}

To further investigate near-Sun kinetic scale intermittency we examine PDFs of increments of \(B_R\) at lags ranging from near energy-containing scales, through the inertial range, down to subproton scales. We first normalize increments (Equation \eqref{eq:inc2}) at each lag by the corresponding standard deviation, and then compute PDFs by calculating the relative frequency of occurrence of increments within designated bins and dividing these frequencies by the bin width to obtain probability densities. The resulting PDFs (Figure \ref{fig:pdf}) are compared
with a Gaussian PDF for reference. Increments at \(\ell = 5000~\di\) measure structures at scales of about half a correlation length, and these non-uniform, ``system-size'' structures exhibit a highly irregular PDF, which nevertheless has the narrowest tails of all. PDFs for the two inertial range lags (100 and \(10~\di)\) show wide, super-Gaussian tails, signifying the presence of outlying ``extreme'' events and intermittency. The \(10~\di\) lag has slightly wider tails, consistent with the well known property of stronger intermittency at smaller inertial-range scales \citep[e.g.,][]{sorriso-valvo1999GRL,chhiber2018MMS}.

\begin{figure}
    \centering
    \includegraphics[width=.44\textwidth]{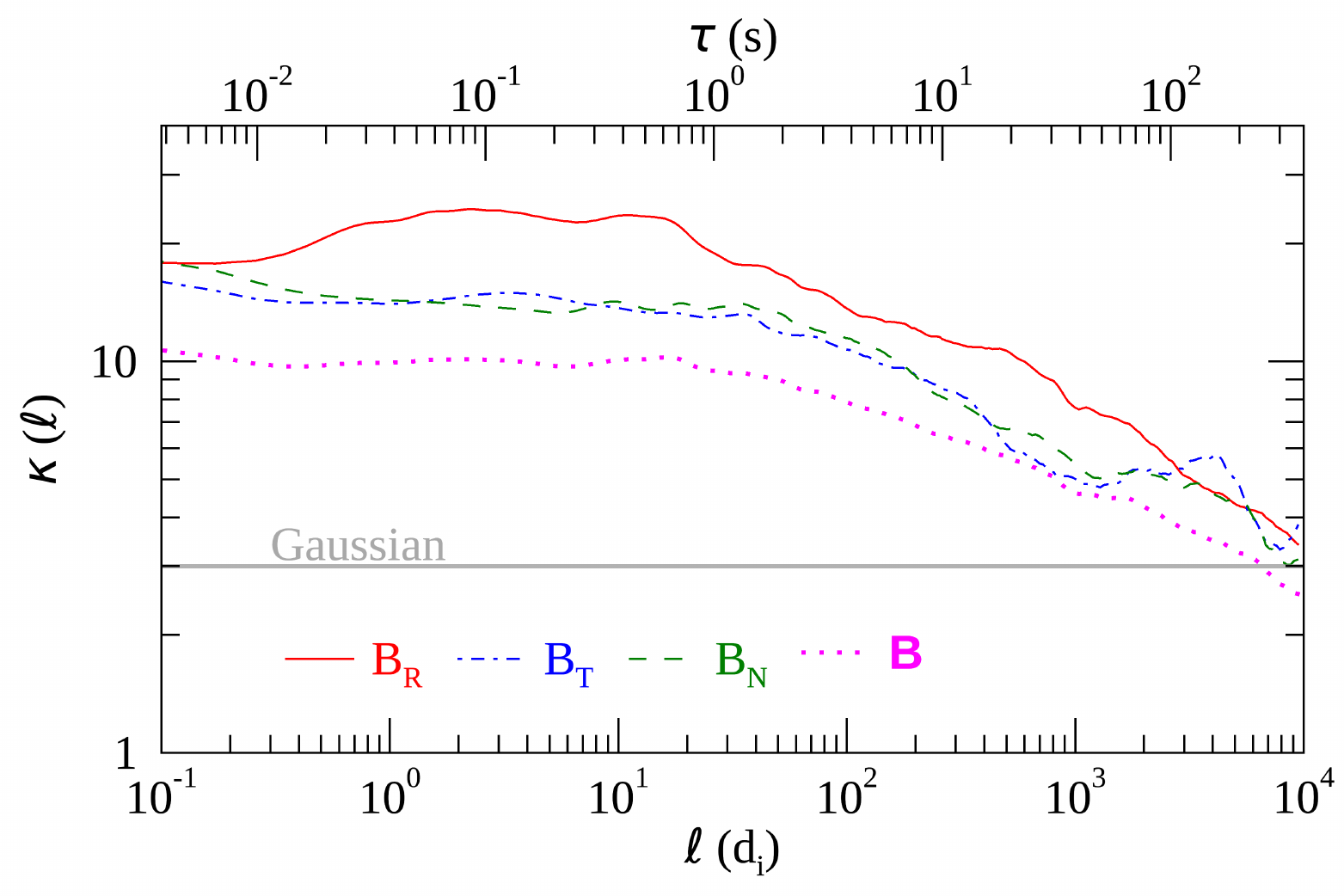}
    \caption{Scale-dependent kurtosis of magnetic field.}
    \label{fig:kurt}
\end{figure}

Moving on to kinetic-range lags (1 and \(0.1~\di\)), we see super-Gaussian tails in PDFs, indicating the continued presence of intermittent structures at these scales. However, the widths of these tails are comparable to (perhaps even slightly narrower than) the \(10~\di\) case, suggesting a saturation of the level of intermittency at proton scales (see also Figure \ref{fig:kurt}, below). Furthermore, the scale similarity suggested by the investigation of scaling exponents in the kinetic range (Figure \ref{fig:struct}) is reaffirmed by the fact that PDFs of the 1 and \(0.1~\di\) lags overlap to large degree. To emphasize this ``monoscaling'', the inset in Figure \ref{fig:pdf} shows PDFs of increments of \(B_R\) for \(\ell = \{0.1,0.3,0.5,0.8,1\}~\di\), \textit{not} normalized by the respective standard deviations as in the main graphic. The outermost (green) curve is for \(\ell=1~\di\) and the innermost (red) curve is for \(\ell = 0.1~\di\). The scale-similar monoscaling of the PDFs is demonstrated by the fact that these PDFs collapse on to each other after being rescaled by their standard deviations \citep[c.f.][]{kiyani2009PRL,osman2015ApJL}. PDFs of \(\delta B_T\) and \(\delta B_N\) behave similarly.

The final diagnostic of intermittency we examine is the SDK, 
a normalized fourth-order moment
that emphasizes the tails of PDFs presented previously:
\begin{equation}
    \kappa (\ell) = \frac{S^{(4)} (\ell)}{\left [S^{(2)} (\ell)\right]^2},\label{eq:kurt}
\end{equation}
where \(S^{(p)}\) can be defined using either Equation \eqref{eq:struc} or \eqref{eq:struc_comp}. \(\kappa (\ell )\) may be thought of as the inverse of the filling fraction for structures at scale \(\ell\); i.e., if \(\kappa (\ell)\) increases with decreasing \(\ell\) then the fraction of volume occupied by structures at scale \(\ell\) decreases with decreasing \(\ell\). The scalar Gaussian distribution has \(\kappa = 3\); a value  \(\kappa > 3\) is a manifestation of wider tails relative to the Gaussian \citep[e.g.,][]{decarlo1997kurtosis}.

Figure \ref{fig:kurt} shows \(\kappa (\ell)\) for individual components of \(\delta \bm{B}\) as well as it's magnitude. All four cases behave similarly -- the kurtosis is near Gaussian at the largest lags, increases to values between 10 and 25 as the lag is decreased across the inertial range to \(\sim 10~\di\), and then stays roughly constant down to \(0.1~\di\). Once again, this indicates a saturation of the intermittency and  scale-similar, monofractal behavior at kinetic scales.\footnote{\label{ftnt:stationarity} To test the robustness of our results and their sensitivity to interval selection (stationarity), the analysis was repeated separately for the first and second halves of the interval, as well as the ``quiet'' period between 02:20 and 02:50 (see Figure \ref{fig:tser}). The results were essentially unchanged, although SDK in the quiet period is relatively smaller and flattens at relatively larger scales (tens of \(\di\)), suggesting weaker multifractality in the inertial range.} This result is consistent with kinetic simulations and \textit{Cluster} observations in the solar wind presented by \cite{wu2013ApJ}. A likely candidate for producing monofractal kinetic-scale kurtosis is a scale-independent fragmentation of current structures between ion and electron scales, as suggested by some kinetic simulations \citep{karimabadi2013coherent}. Note the marked contrast to Figure 8 of \cite{chhiber2018MMS}, where SDK is re-Gaussianized at kinetic scales presumably due to terrestrial foreshock activity and/or instrumental noise in \textit{MMS} measurements.

\section{Discussion and Conclusions}\label{sec:Disc}

In this paper we investigated intermittency in near-Sun solar wind observations of inertial and kinetic range magnetic turbulence, using standard measures including SDK and scaling of higher-order moments up to eighth. Use of a unique \psp\ FIELDS dataset, merged from fluxgate and search-coil magnetometer measurements \citep{bowen2020JGR}, enables study of high frequencies well into the subproton scales, taking Taylor's hypothesis into account. 
Our main results extended several prior studies and clarified outstanding questions concerning solar wind intermittency. First, we observed clearly a monofractal, non-Gaussian, subproton kinetic range, consistent with near-Earth observations \citep{kiyani2009PRL,alberti2019Entropy}. In particular, with \psp\ data close to the sun at 36 $\rs$, far from any foreshock activity, and measurements unaffected by noise \citep{koga2007pre,chian2009AnGeo,wan2012ApJ,wu2013ApJ,chhiber2018MMS}, it is possible to establish clearly that the kurtosis does not re-Gaussianize at sub-ion scales and the statistics remain intermittent. Another major result of interest from the perspective of turbulence theory \citep{benzi1993PRE,benzi1993EPL} is that implementation of ESS 
for sub-ion scales causes a collapse to linear, Kolmogorov-like
behavior for the scaling exponents -- consistent with results reported for kinetic simulation \citep{wan2016PoP}. As far as we are aware this has not been previously demonstrated for magnetic fluctuations in the solar wind. Finally, we report evidence that the magnetic field in near-Sun solar wind exhibits multifractal scaling in the inertial range \citep[c.f.,][]{zhao2020ApJ,alberti2020psp}, which is consistent with near-Earth observations \citep[][and references therein]{bruno2019ESS}, as well as kinetic simulations of turbulence
\citep{leonardis2016pop,wan2016PoP}.
Multifractal inertial-range scaling of higher-order moments is a familiar result in large turbulent MHD systems \citep{politano1998EuroPhysLet,wan2012ApJ}.

We emphasize that the present results are enabled by the the unique orbital position of \psp, along with the high-cadence low-noise character of the 
FIELDS/SCaM magnetic field dataset. Even with the clarifications this analysis provides, there remain unanswered questions. One major outstanding 
issue is why the subproton range becomes monofractal. This implies self-similarity  (or ``rescaling'') of the underlying PDF over that range \citep[e.g.,][]{kiyani2009PRL}.
One possible interpretation is that the range between proton and electron scales is populated by scale-invariant sheet-like concentrations of electric current density. In fact, large numbers of highly dynamic subproton-scale current sheets are seen in kinetic simulations \citep[e.g.,][]{karimabadi2013coherent} and have been inferred in observations \citep[e.g.,][]{retino2007}. This may be distinguished from an effect of incoherent 
linear (noninteracting) waves, which may be expected to produce a return to Gaussianity \citep{koga2007pre,chhiber2018MMS}, and not an onset of monofractal scaling. However, a rigorous connection of structure with monofractality remains to be established and is deferred to future research.  

\acknowledgments 
We thank A. Chasapis for useful discussions. This research was supported in part by the \psp\ mission under the \isois\ project (contract NNN06AA01C) and a subcontract to University of Delaware from Princeton (SUB0000165), and NASA HSR grant 80NSSC18K1648. We acknowledge the \psp\ mission for use of the data, which are publicly available at the \href{https://spdf.gsfc.nasa.gov/}{NASA Space Physics Data Facility}. FIELDS data are publicly available at \url{https://fields.ssl.berkeley.edu/data/}.

\appendix

\numberwithin{equation}{section}

\section{Relating the intermittency parameter to the scaling of kurtosis}\label{sec:app}
Here we briefly examine the relationship between the intermittency parameter \(\mu(p)\) and the scaling behavior of SDK. From Equations \eqref{eq:K62} and \eqref{eq:kurt} we get
\begin{equation}
    \kappa(\ell) =C_\kappa \times  \ell^{\mu(4)-2\mu(2)} = C_\kappa \ell^{-\alpha}, \label{eq:kurt_mu}
\end{equation}
where $C_\kappa$ is a constant and we have defined 
\begin{equation}
\alpha=2\mu(2)-\mu(4).\label{eq:kurt_mu2}
\end{equation}
For homogeneous K41 turbulence we have \(\mu(p)=0\) for all \(p\), and \(\kappa (\ell)\) is constant. For K62 turbulence the intermittency parameters $\mu(*)$ may be nonzero, and, within a range of lags defined by either monofractality or multifractality, $C_\kappa$ does not vary with cascade rate or $\ell$ but may depend (in hydrodynamics) on the local Reynolds number \citep{kolmogorov1962JFM}. For monofractal behavior \(\zeta(p)=p/3+\mu(p)\) is linear in \(p\), and therefore \(\mu(4) = 2\mu(2)\), which implies a scale-invariant \(\kappa\), i.e., \(\kappa\) is constant function of \(\ell\). Indeed, from the middle panel of Figure \ref{fig:struct} we find for the monofractal kinetic range (blue curve) that \(\mu(2)= 1.73\) and \(\mu(4)= 3.45\), so that \(\mu(4)= 1.99 \mu(2)\). Finally, for multifractal statistics one expects \(|\mu(4)| > |\mu(2)|\), but if \(\kappa\) is a decreasing function of \(\ell\) (as is generally observed) then \(\alpha\) remains positive, and the observed exponent \(\alpha\) constrains \(\mu(2)\) and \(\mu(4)\). From Figure \ref{fig:kurt} we have \(\alpha\approx 0.17\) for the pink curve in the inertial range (\(\ell=20\) - \(10^3~\di\)); from the middle panel of Figure \ref{fig:struct}, we find for the inertial range (red curve) that \(2\mu(2) - \mu(4)= 0.17\). Therefore the relationship between the scaling behavior of the SDK and the intermittency parameter is well described by Equations \eqref{eq:kurt_mu} and \eqref{eq:kurt_mu2}.

\bibliography{chhibref}
\listofchanges
\end{document}